\documentclass[11pt,a4paper]{article}
\pdfoutput=1
\usepackage{amssymb,amsmath,amsfonts,mathrsfs}
\usepackage{amsmath}
\usepackage{slashed}
\usepackage{jheppub}
\usepackage{mathrsfs}
\usepackage{latexsym}
\usepackage{multirow}
\usepackage[usenames,dvipsnames,table]{xcolor}
\usepackage{graphicx}
\usepackage{epsfig}  
\usepackage{dcolumn}
\usepackage{bm}
\usepackage{dcolumn}
\usepackage{textcomp}
\usepackage{float}
\usepackage{subfig}
\usepackage{hypcap}
\usepackage[]{hyperref}
\usepackage{makecell}
\usepackage{color}
\usepackage{pifont}
\usepackage{appendix}
\usepackage{url}
\usepackage{cancel}
\allowdisplaybreaks[1]
\usepackage{slashed}

\usepackage[normalem]{ulem}
\definecolor{darkgreen}{cmyk}{1,0,1,0.4}

\definecolor{pink}{cmyk}{0.4,1,0.3,0}

\def\com2#1{\textcolor{red}{\it{#1}}}
\hypersetup{
  bookmarks=true,         
  unicode=false,          
  pdftoolbar=true,        
 pdfmenubar=true,        
 pdffitwindow=true,     
 pdfstartview={FitH},    
 pdfsubject={GUT and Topological defects},   
 pdfnewwindow=true,      
 pdfcreator={RevTeX},
 colorlinks=true,       
 linkcolor=red,          
 citecolor=blue,        
 filecolor=black,      
 urlcolor=blue,           
  }

\makeatletter
\renewcommand{\fnum@table}{\textbf{\tablename~\thetable}}
\renewcommand{\fnum@figure}{\textbf{\figurename~\thefigure}}
\makeatother
\title
{MeV scale model of SIMP dark matter, neutrino mass and leptogenesis }

\author[a]{Subhendra Mohanty,}  
\author[a]{Ayon Patra,}
\author[a]{Tripurari Srivastava}  
   \affiliation[a]{Theoretical Physics Division, Physical Research Laboratory, Ahmedabad-380009, India}  

\emailAdd{mohanty@prl.res.in}
\emailAdd{ayon@okstate.edu}
\emailAdd{tripurari@prl.res.in}
\abstract
{We consider a simple extension of the Standard Model with two singlet scalar fields and three heavy right-handed neutrinos. One of the scalar fields serves as an MeV scale dark matter and its stability is ensured by the introduction of an extra $Z_2$ symmetry. The second scalar (which is even under the $Z_2$ symmetry) generates the mass term of the scalar, contributes to the $3 \rightarrow 2$ annihilation process required for the correct relic density of the dark matter and it also contributes to the leptogenesis. The right-handed neutrinos are responsible for the generation of light neutrino masses through Type-I seesaw mechanism. The decay of the heavy right-handed neutrino can generate the lepton asymmetry which can then be converted to baryon asymmetry through sphaleron transitions.} 

\begin{document}
\maketitle
\flushbottom
\newpage
\section{Introduction}
The Weakly interacting massive particle (WIMP)~\cite{Kolb:1990vq,Roszkowski:2017nbc} paradigm of dark matter (DM) is severely constrained due to the non-observation of any signal in DM direct detection experiments~\cite{Kim:2012rza,Aprile:2015uzo,Aprile:2017ngb}. This calls for an extension of the DM parameter space to a lower mass range which is not very well probed in direct detection experiments. The amount of dark matter in the universe is also well measured from various observations, e.g., CMB~\cite{Aghanim:2018eyx} and large scale structure surveys~\cite{Refregier:2003ct}. At the scale of galaxies, the cold dark matter model is in good agreement with the observation. However there are certain problems like lesser observation of dwarf galaxies (missing satellite problem)~\cite{Moore:1999nt,Klypin:1999uc} and the absence of very bright galaxies (too big too fail problem)~\cite{10.1111/j.1745-3933.2011.01074.x,10.1111/j.1365-2966.2012.20695.x} which are both predicted from N-body simulations. Simulations also predict that in the galactic distribution of dark matter, there is a large concentration at the core which is again not observed in rotation curves of galaxies (core cusp problem)~\cite{Burkert:1995yz,Salucci:2007tm}. These problems of $\Lambda$CDM model can be solved by introducing large self-interactions in the dark matter~\cite{PhysRevLett.84.3760,Balberg:2002ue,Tulin:2017ara}. 

The observed relic density, which in the WIMP paradigm, could be explained by decoupling of the $2\rightarrow 2$ annihilation no longer works for MeV scale DM. On the other hand, the abundance of an MeV scale DM naturally satisfies the relic density constraints if there exists a $3\rightarrow 2$ mechanism for DM annihilation~\cite{Hochberg:2014kqa,Hochberg:2015vrg,Bernal:2015bla,Choi:2015bya,Dey:2016qgf,Cline:2017tka,Kamada:2017tsq,Choi:2017zww,Ho:2017fte,Heikinheimo:2018esa,Hochberg:2018vdo,Hochberg:2018rjs,Namjoo:2018oyn,Dey:2018yjt,Choi:2019zeb,Bhattacharya:2019mmy}. This requires a large self-interaction among the DM particles and is popularly known as the strongly interacting massive particle (SIMP) DM scenario \cite{Bernal:2015xba,Choi:2016hid,Choi:2016tkj,Bernal:2017mqb,Chauhan:2017eck}. In the usual $3 \rightarrow 2$ mechanism, three DM particles can annihilate into two DM particles in the final state. In our model, we have invoked a $Z_2$ symmetry to ensure the stability of DM ($\phi$), which however prevents this 3$\phi$ to 2$\phi$ annihilation process. We therefore introduce a second scalar ($\delta$) which assists in obtaining the correct relic density via a $\phi \phi \phi \rightarrow \phi \delta$ annihilation process. We write the most general potential involving the scalars $\phi$ and $\delta$ which respects the $Z_2$-odd nature of $\phi$. In this potential, by giving $\delta$ a vacuum expectation value, we generate a mass of the $\phi$ and the $\delta$ in the $\sim 10$ MeV scale. Although the scalar $\delta$ does not contribute to the relic density, it can interact with the right-handed neutrinos to generate the observed baryon asymmetry of the universe through low scale leptogenesis \cite{Fukugita:1986hr,Pilaftsis:1997jf,Pilaftsis:2003gt,Kang:2006sn,Kang:2014mea}. In this model we have introduced three right-handed neutrinos (two of them with masses around 10 TeV and one much lighter with mass around 10 MeV) to obtain the light neutrino masses by Type-I seesaw mechanism \cite{Minkowski:1977sc,Yanagida:1979as,Sawada:1979dis,Levy:1980ws,VanNieuwenhuizen:1979hm,Mohapatra:1979ia} while also being able to generate the required lepton asymmetry. The decay process of the heavy right-handed neutrino can produce the lepton asymmetry through the interference of the tree-level and the one-loop diagram involving the lighter right-handed neutrino and the scalar $\delta$. 

The rest of the paper is organized as follows. In Sec.~\ref{model} we introduce the model and present a detailed analysis of the scalar sector. The calculation of the DM relic density is performed in Sec.~\ref{RD} where we scan over the allowed parameters which provide the correct experimentally observed relic abundance. In Sec.~\ref{SI} we look at the constraints on the self-interaction cross-section for a SIMP DM arising from various astrophysical observations and study their effect on our allowed parameter space. A detailed analysis of the neutrino sector is performed in Sec.~\ref{NM}. Our scan results show that a clear hierarchy among the neutrino Dirac masses is required to satisfy the experimentally observed neutrino mass-squared differences and mixings. In Sec.~\ref{LEP} we study the mechanism to achieve low scale leptogenesis in our framework. Sec.~\ref{N2BD} includes a discussion on the possible constraints arising from neutrino-less double beta decay processes. We present our conclusions in Sec.~\ref{CON}.

\section{Model} \label{model}

We extend the SM framework with two singlet scalars and three right-handed neutrinos. The gauge group of the SM remains unchanged but we need to introduce a discrete $Z_2$ symmetry to ensure the stability of the dark matter. The particle spectrum of the model is given in Table~\ref{tab:one}. 
\begin{table}
\begin{center}
{\renewcommand{\arraystretch}{1.5}
\begin{tabular}{c|c|c|c|c} \hline
Field & $SU(3)_C$ & $SU(2)_L$ & $U(1)_Y$ & ${Z}_2$ \\ \hline
$Q_{L_i} = \left (\begin{array}{c} u\\ d \end{array} \right )_{L_i}$ & 3 & 2 & $\frac{1}{3}$ & + \\
$u_{R_i}$ & 3 & 1 & $\frac{4}{3}$ & + \\ 
$d_{R_i}$ & 3 & 1 & $-\frac{2}{3}$ & + \\ 
$l_{L_i} = \left (\begin{array}{c} \nu \\ e \end{array} \right )_{L_i}$ & 1 & 2 & -1 & + \\ 
$e_{R_i}$ & 1 & 1 & -2 & + \\ 
$N_i$ & 1 & 1 & 0 & + \\ 
$H= \left (\begin{array}{c} H^+ \\H_0 \end{array} \right )$ & 1 & 2 & 1 & + \\
$\phi$ & 1 & 1 & 0 & - \\ 
$\delta$ & 1 & 1 & 0 & + \\ 
\end{tabular}}
\caption{Particle spectrum for the Model. Here $i$ is the generation index.}
\label{tab:one}
\end{center}
\end{table}
The quarks and leptons remain the same as in the SM but we introduce three right-handed neutrinos ($N_i$) which help in generating the light neutrino masses and mixings via Type-I seesaw mechanism. Similar to the SM, the Higgs doublet $H$ is responsible for the electroweak symmetry breaking and generation of quark and lepton masses. As can be seen from Table~\ref{tab:one}, the scalar singlet $\phi$ is odd under the $Z_2$ symmetry while all other particles are even. The scalar $\phi$ thus cannot decay and becomes the DM candidate in this model. The other scalar singlet $\delta$ can acquire a non zero vacuum expectation value (VEV) and generate the mass for the DM. The scalar $\delta$ actually has multiple roles in this model -- in addition to contributing to the DM mass, it helps in obtaining the correct DM relic density by assisting in the $3 \to 2$ DM annihilation process and also plays an important role in generating the lepton asymmetry. One of the heavy right-handed neutrinos can decay into a light fermion and a scalar boson directly through a tree level process but this cannot produce any Charge-Parity (CP) asymmetry. The one-loop corrections to this process includes vertex correction and self-energy corrections, and their interference with the tree-level decay process give rise to the required CP asymmetry. This will be discussed in more details in section~\ref{LEP}. 

The non-zero vacuum expectation values for the scalar fields are 
\begin{equation}
\left< H \right> = v_H,~~\left< \delta \right> = v_\delta,
\end{equation}
while the scalar potential in the model is given as:
\begin{eqnarray}
\mathcal{L} & = & \frac{\lambda_{11}}{4!} \phi^4 + \frac{\mu_{\phi}^2}{2} \phi^2 +\frac{\lambda_{12}}{4} \phi^2 \delta^2 
+ \frac{\mu_{\delta}^2}{2} \delta^2 + \frac{\mu_{22}}{3!} \delta^3 +\frac{\lambda_{22}}{4!} \delta^4 +\frac{\mu_{21}}{2} \phi^2 \delta \notag \\
&+& \frac{\lambda_{13}}{2} \phi^2 H^\dagger H + \mu_{23} \delta H^\dagger H +  \frac{\lambda_{23}}{2} \delta^2 H^\dagger H + \mu_H^2 H^\dagger H + \lambda_{33} \left[ H^\dagger H\right]^2.
\end{eqnarray} 

Measurement of the Higgs invisible decay branching ratio puts stringent constraints on some of these couplings. Recent results from the LHC gives an upper limit on the invisible decay of the Higgs boson as BR$(H\to \rm{invisible}) < 0.19$~\cite{Sirunyan:2018owy}. In our model, the process contributing to it comes from the $H\to \phi \phi$ channel. Calculating the decay width of this channel gives:
\begin{equation}
\Gamma (H\to\phi \phi)= \frac{\lambda_{13}^2 v_{H}^2}{32 \pi m_H} \Big(1-\frac{4 m_\phi^2}{m_H^2}\Big)^{\frac{1}{2}}.
\end{equation}
The numerical value of $\lambda_{13}< \mathcal{O}(10^{-2})$ ensures the Higgs invisible decay constraint are satisfied. 

The scalar field $\phi$, being odd under $Z_2$, does not mix with the other fields and becomes the DM candidate with its mass given as
\begin{equation}
M_\phi^2 =  \left( \frac{\lambda_{12}}{2} v_\delta^2 + 2\lambda_{13} v_H^2 + \mu_{12} v_\delta +2 \mu_\phi^2 \right).
\end{equation} 
In order to obtain the correct relic density, $M_\phi$ must be in the MeV range with $\lambda_{12} \sim 1$. Hence we choose $v_\delta$ to be of order MeV as well. The value of $v_H$, on the other hand, is required to be $v_H \equiv v_{EW} =174$ GeV. This forces $\lambda_{13}$ to be extremely small ($\lesssim 10^{-6}$) as otherwise a large fine-tuning will be required to obtain the low DM mass. 

The tiny $\lambda_{13}$ coupling also plays an important role in evading the DM direct detection experimental constraints. Current dark matter direct detection experiments are based on nucleon recoil energy measurement from their scattering with the DM~\cite{Kang:2018odb}. However, for sub GeV dark matter, these bounds are quite weak although recent experiments, e.g., CRESST-III~\cite{Petricca:2017zdp}, Xenon1T~\cite{Aprile:2018dbl} and CDEX-1B~\cite{Liu:2019kzq} etc. provide limit for some range in sub GeV DM. The small $\lambda_{13}$ coupling results in the DM scattering with the quarks (via a Higgs boson) to be highly suppressed compared to the limit from direct detection experiments. There are other experiments, e.g., Xenon10~\cite{Essig:2017kqs}, super CDMS~\cite{Agnese:2018gze} and SENSEI~\cite{Crisler:2018gci} that use DM electron scattering to also provide constraints over cross-section. Since the Higgs couplings to the DM particle and to the electrons are both extremely small, we can easily satisfy the experimental bounds in this case as well.

The neutral component of the doublet scalar $H_0$ and the singlet $\delta$ fields mix to form a $2\times 2$ mass-squared matrix given as
\begin{equation}
M_H^2 = \begin{bmatrix}
 \frac{1}{2} \lambda_{22} v_\delta^2 + 2\lambda_{23} v_H^2 +6 \mu_{22} v_\delta + 2\mu_\delta^2 & -\frac{v_\delta} {3\sqrt{2}v_H} \left( \lambda _{22} v_\delta^2 + 18 \mu_{22} v  _\delta + 12\mu_\delta^2 \right) \\ -\frac{v_\delta} {3\sqrt{2}v_H} \left( \lambda _{22} v_\delta^2 + 18 \mu_{22} v  _\delta + 12\mu_\delta^2 \right) & 4\lambda_{33} v_H^2
\end{bmatrix}.
\end{equation} 
As can clearly be seen, the mixing between the $H_0$ and the $\delta$ field is proportional to $\frac{v_\delta}{v_H}$ and hence is extremely small.  

\section{Relic abundance} \label{RD}
The SM Higgs mixings with the singlet scalars in our model remain very small as was discussed in Section.~\ref{model}. The additional scalars $\phi$ and $\delta$ thus remain largely secluded. The relic abundance of dark matter can be computed via annihilation of dark matter $\phi$ through $3\to 2$ process. Choosing $m_\phi< m_\delta< 2m_\phi$ and imposing the discrete $Z_2$ symmetry, the dominant contribution for the annihilation of $\phi$ is through $\phi\phi\phi \to \phi \delta$. In addition to this, the $\delta \delta \to \phi \phi$ process also needs to be included to compute the evolution of number density of $\phi$. The Feynman diagrams for the relevant processes contributing to the relic abundance are given in the Fig.~\ref{ppppd_relic}.

\begin{figure}[h]
\begin{center}
\includegraphics[width=0.45\textwidth]{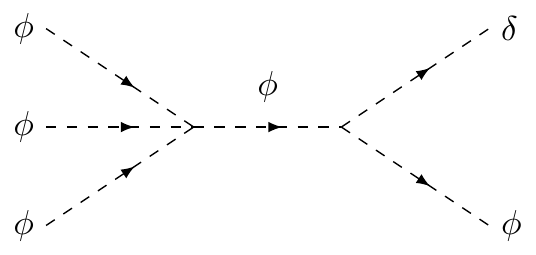} 
\includegraphics[width=0.25\textwidth]{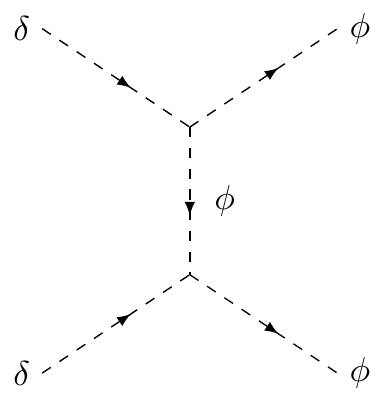} 
\caption{\label{ppppd_relic} Feynman diagram for processes contributing to the calculation of relic density.}
 \end{center}
\end{figure}

Evolution of number density of dark matter is governed by the set of Boltzmann's equations given by:
\begin{eqnarray}
\frac{d n_\phi}{dt}+3H n_\phi&=& -\frac{1}{3} \left<\sigma v^2 \right>_{3\to2} (n_\phi^3-n_\phi n_\delta n_\phi^{eq})+ \left< \sigma v \right>_{2\to2} (n_\phi^2- {n_\delta^{2}})\\
\frac{d n_\delta}{dt}+3H n_\delta&=& +\frac{1}{6} \left<\sigma v^2 \right>_{3\to2} (n_\phi^3-n_\phi n_\delta n_\phi^{eq})- \left<\sigma v \right>_{2\to2} (n_\phi^2- {n_\delta^{2}}).
\end{eqnarray}
Here the number densities are defined as:
\begin{equation}
n_i=g_i \int \frac{ d^3 p_i}{(2\pi)^3} f_i,
\end{equation}
where $f_i$ is the distribution function for i-th particle. The factors in front of thermal average cross-sections are to avoid double counting (see~\cite{Yang:2019bvg}).
We can perform a change of variables to solve the equations by assuming $Y_i=n_i/s$ and $x=m_\phi/T$ where $s$ is the entropy density given as $(2\pi^2 g^{*}T^3/45)$ and $T$ is the temperature. The number density evolution equation can be written as:
\begin{eqnarray}
\frac{d Y_\phi}{dx}&=& -\frac{ x s^2}{3H(m_\phi)} \left< \sigma v^2 \right>_{3\to2} (Y_\phi^3-Y_\phi Y_\delta Y_{\phi}^{eq})+\frac{xs}{H(m_\phi)} \left<\sigma v \right>_{2\to2} (Y_\phi^2- {Y_\delta^{2}})\\
\frac{d Y_\delta}{dx}&=& +\frac{x s^2}{6H(m_\phi)} \left< \sigma v^2 \right>_{3\to2} (Y_\phi^3-Y_\phi Y_\delta Y_{\phi}^{eq})-\frac{xs}{H(m_\phi)} \left<\sigma v \right>_{2\to2} (Y_\phi^2- {Y_\delta^{2}}),
\end{eqnarray}
where $H(m_\phi)=\sqrt{\frac{\pi ^2 g^*}{ 90}} \frac{m_\phi^2}{{M_{pl}}}$.

Thermal average cross-section can be computed from the model as follows:
\begin{eqnarray}
\langle \sigma_{3\to2} v^2 \rangle &=& \frac{1}{ n_\phi^3}\int d\Pi_\phi d\Pi_\phi d\Pi_\phi d\Pi_\delta d\Pi_\phi (2\pi)^4 \\&& \delta^4 (p_1(\phi) + p_2(\phi)+p_3 (\phi)-p_4 (p_\delta)-p_5 (\phi)) f_ \phi^3 |\mathcal{M}_{\phi\phi\phi\to \delta \delta}|^2,
\end{eqnarray}
where $\Pi_i=\frac{d^3 p_i}{(2\pi)^3 2 E_i} $.

For the $3\to 2$ process, the thermal average cross-section can be written as~\cite{Berlin:2016gtr}:
\begin{eqnarray}
\left< \sigma_{3\to2} v^2 \right>&=& \frac{\lambda_{11}^2\mu_{\rm eff}^2}{64 \pi m_\phi^3}\sqrt{\Big(1-\frac{(m_\phi+m_\delta)^2}{9m_\phi^2}\Big)\Big(1-\frac{(m_\delta-m_\phi)^2}{9m_\phi^2}\Big)}\frac{1}{64m_\phi^4},
\end{eqnarray}
where $\mu_{\rm eff} = \mu_{12} + \frac{\lambda_{12}}{2} v_\delta$. Here, we have assumed $\phi$ to be non-relativistic. We solve the coupled Boltzmann equations numerically to evaluate the yield $Y_\phi$ at freeze out temperature. The relic abundance can be obtained from it and is given by:
\begin{equation}
\Omega_\phi h^2\approx 2.755\times 10^8 m_\phi Y_\phi(T_f) 
\end{equation}
We show the freezeout of $\phi$ in Fig.~\ref{ppppd_relic} by plotting the evolution of $Y_\phi$ with  $x~(=m_\phi/T)$, where $T$ is temperature. Here we take the masses $m_\phi=40$ MeV, $m_\delta=60$ MeV, and coupling $\lambda_{11}=4$. The black solid line in the figure represents the value of $Y_\phi$ while the dotted red line is the Maxwell-Boltzmann equilibrium distribution function.
\begin{figure}[h!]
\begin{center}
\includegraphics[width=0.75\textwidth]{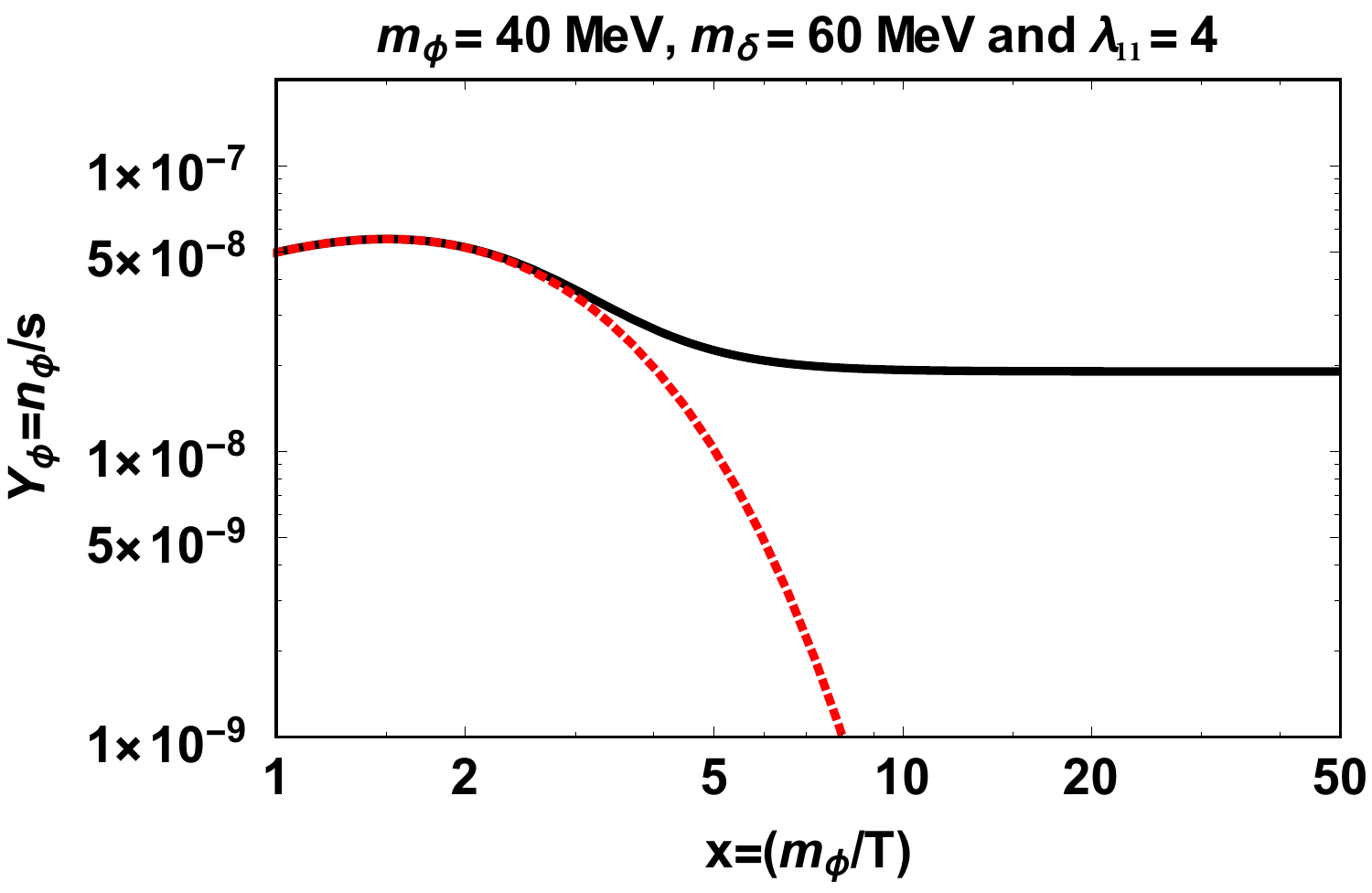} 
\caption{\label{ppppd_relic} Temperature evolution of number density of $\phi$ with $v_\delta$ = 50 MeV. Black solid line represents the yield $Y_\phi$ evolution with dimension less variable $x$. Red dashed line shows the variation of $Y_\phi^{eq}$.}
 \end{center}
\end{figure}

\begin{figure}[h!]
\begin{center}
\includegraphics[width=0.65\textwidth]{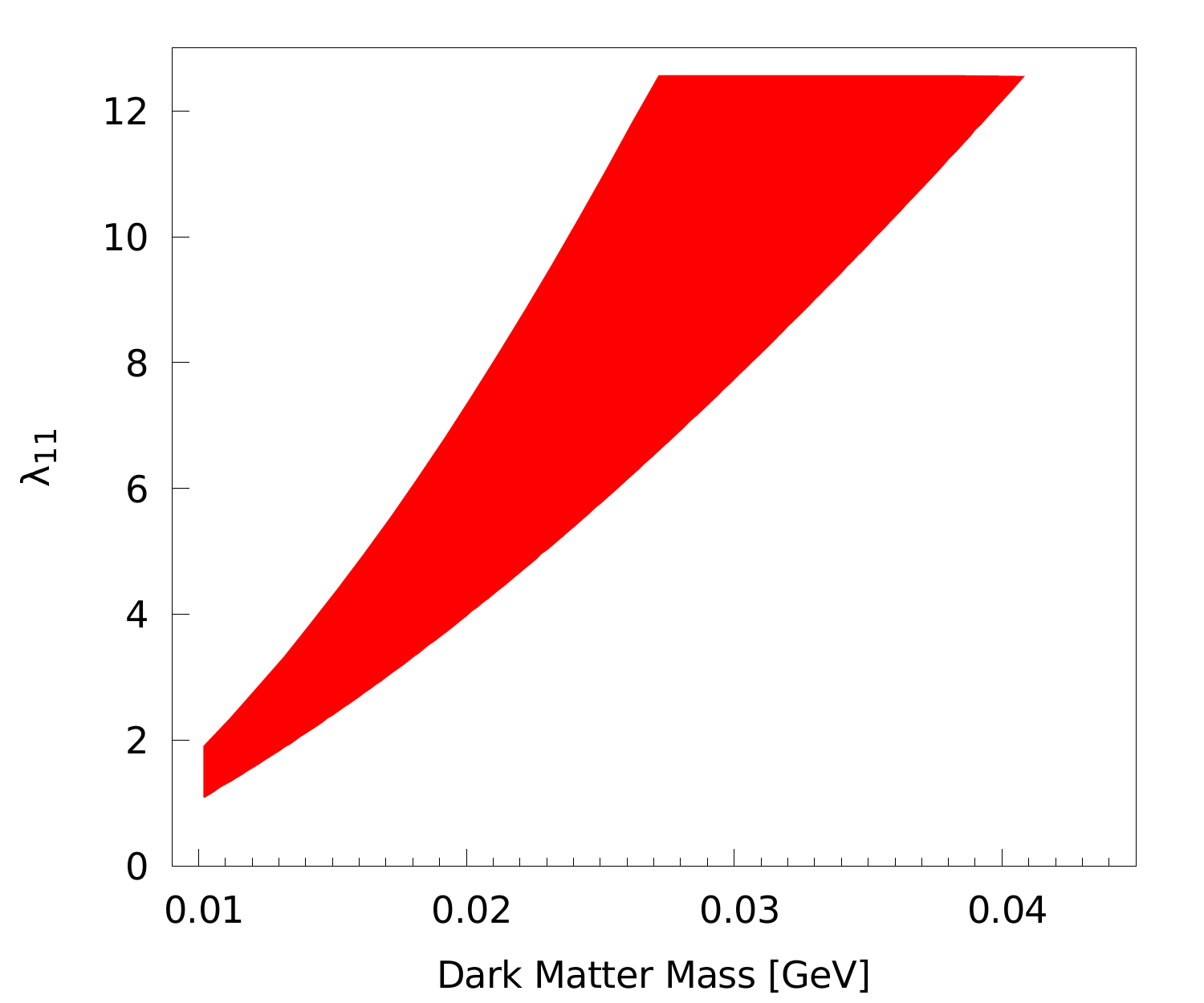} 
\caption{\label{region_relic} Correlation of coupling and mass of DM that satisfy the relic abundance from PLANCK.}
 \end{center}
\end{figure}

We analyze the relic abundance of DM to obtain the parameters that satisfy the PLANCK data. The recent PLANCK observation limit on DM relic abundance is given as~\cite{Aghanim:2018eyx}:
\begin{equation*}
\Omega h^2=0.1206\pm 0.0021.
\end{equation*}
We analyze the model in the light of the constraints from the Planck data for DM relic density. To have annihilation via mentioned channel, we choose $m_\phi<m_\delta<2m_\phi$ and randomly vary the values of $\lambda_{11}$ and $\lambda_{12}$ in their perturbative range of $[0,4\pi]$. In Fig.~\ref{region_relic}, we show the region of parameter space that satisfy the $1\sigma$ range of Planck data in the $M_\phi - \lambda_{11}$ plane. The plot reveals that as the DM mass increases, a larger value of $\lambda_{11}$ is required to satisfy the relic density constraint. This is quite expected as the annihilation cross-section decreases with an increase in the DM mass and it increases with an increase in the coupling. Thus, simply increasing the DM mass itself would result in an overabundance of DM density and the value of $\lambda_{11}$ should also increase to satisfy the observed bound.

Next, we study the effect of changing the mass difference between the DM and the singlet scalar field. 
\begin{figure}[h!]
\begin{center}
\includegraphics[width=0.85\textwidth]{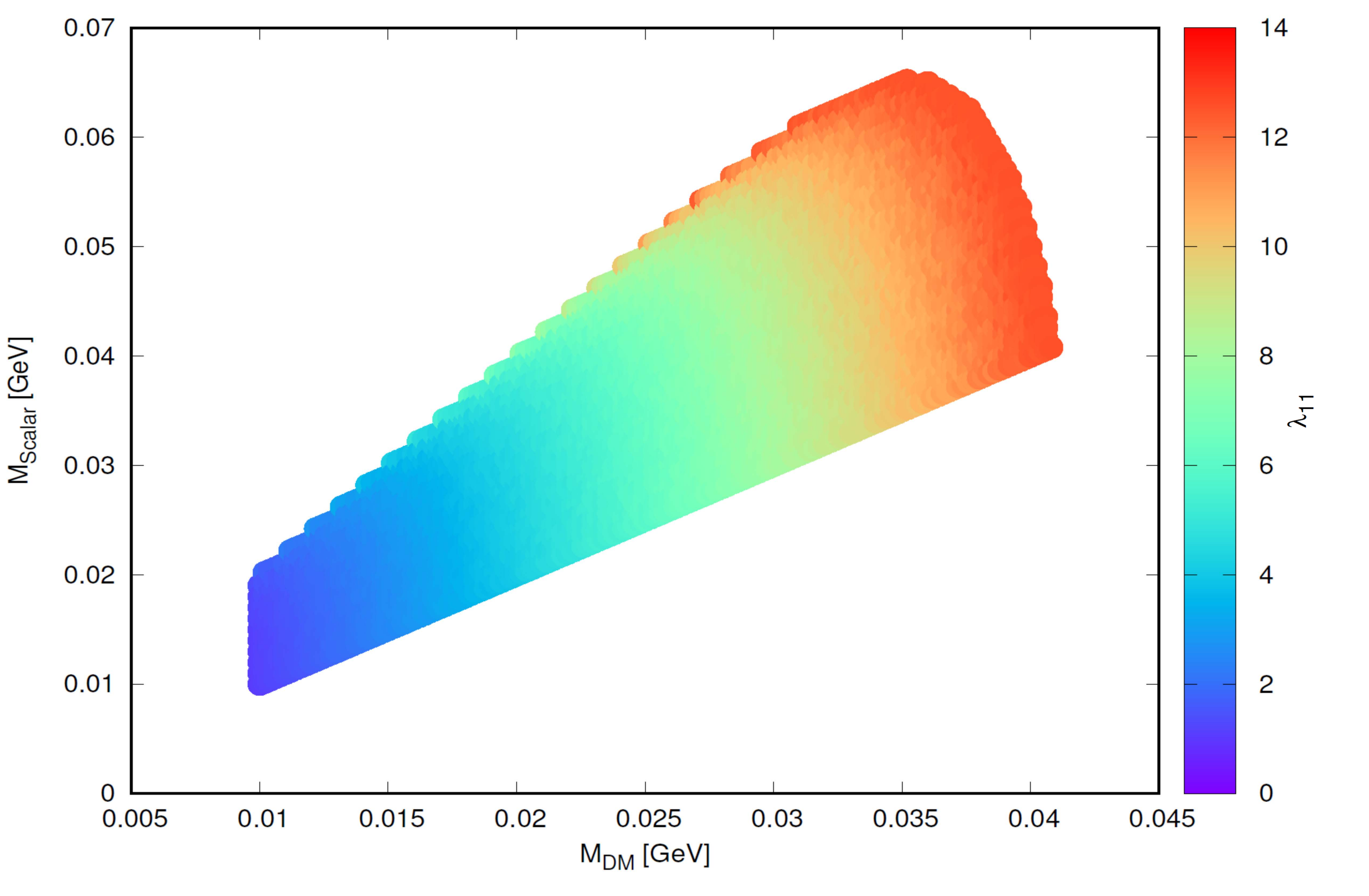} 
\caption{\label{fig:mdmmphi} Allowed values of scalar mass which satisfy the relic density constraint for DM mass varying from 10 MeV to 100 MeV.}
 \end{center}
\end{figure}
In Fig.~\ref{fig:mdmmphi} we plot the scalar boson mass with respect to the DM mass. For a given dark matter mass $m_\phi$, the scalar boson mass $M_\delta$ is varied between $M_\phi<M_\delta<2 M_\phi$. The plot also represents the magnitude of the self-coupling ($\lambda_{11}$) required for each point through a color coding. Similar to the previous plot, an increase in the DM mass must be accompanied by an increase in the self coupling in order to satisfy the relic density constraints. It is also evident from the plot that for a fixed DM mass, the required value of $\lambda_{11}$ increases as $M_\delta$ increases. This is again quite natural since an increase in the scalar boson mass results in a reduction of the phase space factor as the mass difference between the initial and final states get smaller. Hence a larger value of the coupling would be required to satisfy the observed relic density. 

\begin{figure}[h!]
\begin{center}
\includegraphics[trim=000 000 000 010,clip,width=0.85\textwidth]{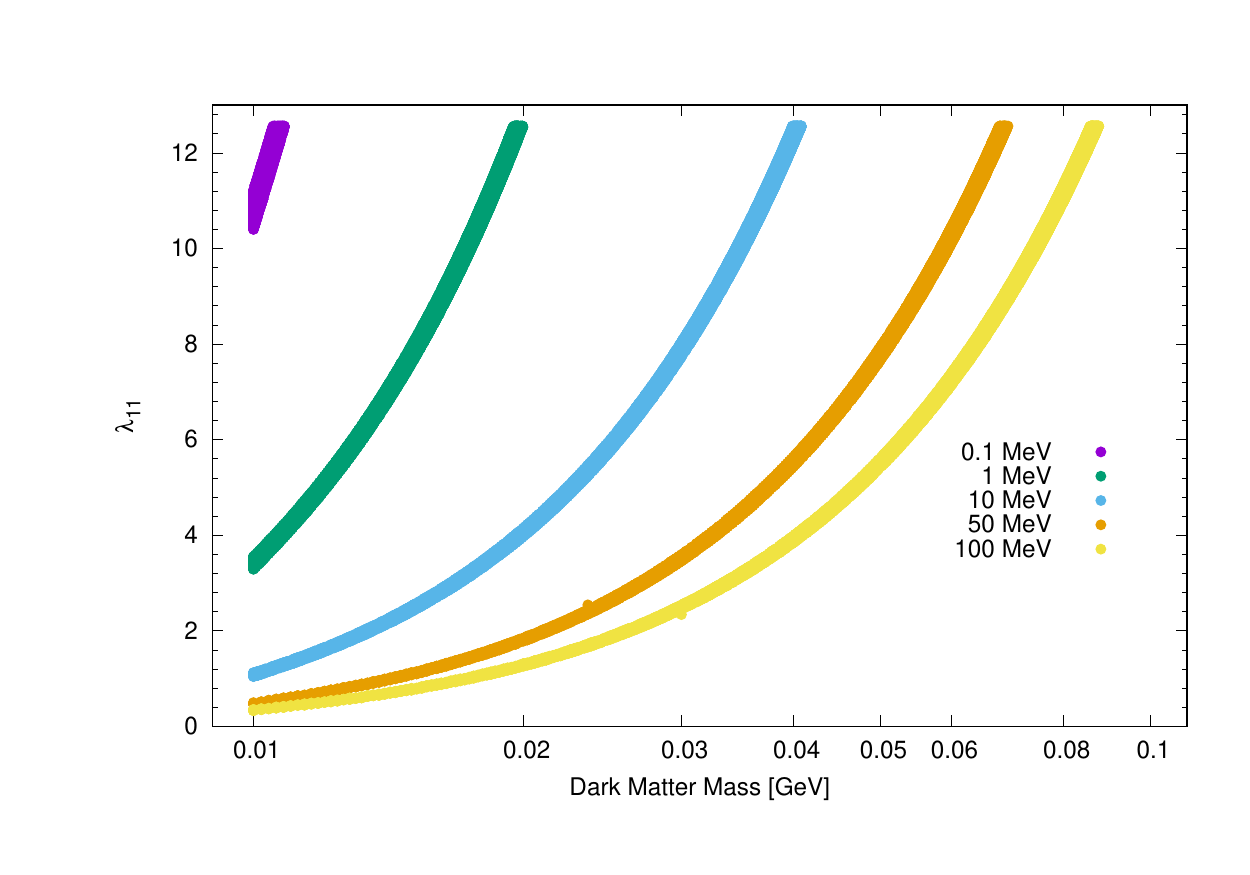} 
\caption{\label{fig:relic_vev} Points satisfying the DM relic density for different values of $v_\delta$. The purple points correspond to $v_\delta = 0.1$ MeV, green for $v_\delta=1$ MeV, blue for $v_\delta=10$ MeV, orange points represent $v_\delta=50$ MeV and yellow points are for $v_\delta=100$ MeV.}
 \end{center}
\end{figure}
Finally we focus on the effect of varying the VEV of the scalar field on the DM relic density. Fig.~\ref{fig:relic_vev} shows five different plots corresponding to five different values of $v_\delta$ and their corresponding allowed parameter points in the $M_\phi - \lambda_{11}$ plane. Here we have chosen the singlet scalar mass to be degenerate with the DM mass and $\lambda_{11} = \lambda_{12}$. The purple points corresponding to the $v_\delta=0.1$ MeV can only satisfy the relic density constraints for large values of the coupling. Since the $\delta \phi \phi$ coupling is proportional to the $v_\delta$, a small value of the VEV requires a large value of the quartic coupling in order to produce enough annihilation of the DM particle. Similarly as the VEV increases one can satisfy the relic density constraints for smaller values of the quartic couplings as well. Another interesting feature of Fig.~\ref{fig:relic_vev} is that the allowed range of DM masses increases as the value of the VEV increases. A quick glance at the $3 \rightarrow 2$ scattering process reveals that its cross-section is proportional to $v_\delta^2$ and hence a larger value of $v_\delta$ will naturally allow $m_\phi$ to be large as well.

\section{Self interaction} \label{SI}

Self-interacting DM can solve many of the discrepancies between experimental observations and N-body simulations~\cite{PhysRevLett.84.3760,Balberg:2002ue}. As an example, numerical simulations in $\Lambda$CDM framework predict the DM density radially diverges at the core of galaxy clusters while actual observations suggest a flat DM distribution \cite{Burkert:1995yz,Salucci:2007tm}. Self-interaction among the DM particles allow them to be scattered by each other and can result in a uniform DM density at the core as is expected from experimental observations. Additionally, observations from galaxy clusters can also put severe constraints on the self-scattering cross-section of DM particles. The mean free path for scattering of DM can be given as:
\begin{equation}
\lambda_{\rm scatt}=\frac{\sigma_{\rm self}}{m_\phi} \rho\,v,
\end{equation} 
where $\sigma_{\rm self}$, $\rho$, $v$ and $m_{\phi}$ are the self-scattering cross-section, density, velocity and mass of dark matter respectively. Observations from galaxy clusters provide the density at the core and a typical velocity $\sim$ 50 $\rm{km s^{-1}}$, while the value of $\lambda_{\rm scatt}$ is around the size of the galaxy cluster. Knowing all these values from astrophysical observations, one can easily provide constraints on the ratio of the self-scattering cross-section to the mass of the DM particle. The constraints on $\sigma_{\rm self}/m_\phi$ from different observations are as follows:
\begin{table}[H]
\begin{center}
    \begin{tabular}{| l | l |}
    \hline
    $\sigma_{\rm self}/m_\phi(\rm{ cm^2/g})$ & Observations  \\ \hline
   $ \sim (1.7\pm 0.7)\times 10^{-4}$ &  Bright cluster galaxies in the 10 kpc core of Abell 3827 ~\text{\cite{Massey:2015dkw}}\\ \hline
    $ \sim 0.1$ & Cores in clusters ~\text{\cite{Elbert_2018,PhysRevLett.116.041302}}\\ \hline
   $ \sim 1.5$ & Abell 3827 subhalos ~\text{\cite{Kahlhoefer:2015vua}}\\ \hline
       $ \sim  1 $ & Abell 520 cluster~\text{\cite{refId0,Jee_2014,Kahlhoefer:2013dca}} \\ \hline
 $\lesssim 1$  & Halo shapes and Bullet cluster~\text{\cite{Randall:2007ph,Peter:2012jh}} \\ \hline
 \end{tabular}
\caption{{\label{tab:sigmam}}Constraints on DM self-scattering cross-section from various observations.}
\end{center} 
\end{table}
 
\begin{figure}[h!]
\begin{center}
\includegraphics[width=0.20\textwidth]{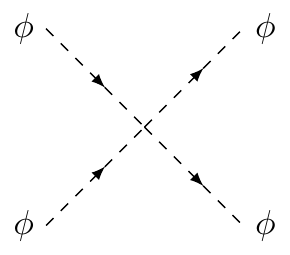} 
\includegraphics[width=0.32\textwidth]{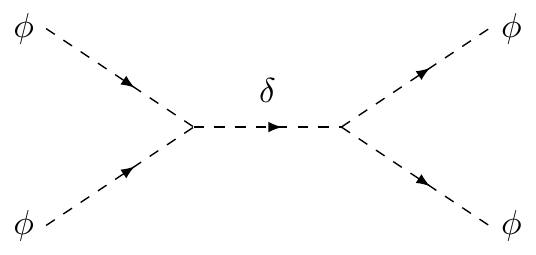}
\includegraphics[width=0.20\textwidth]{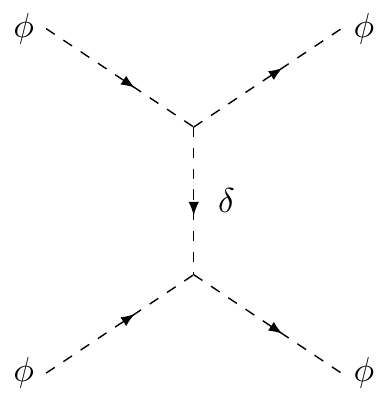} 
\includegraphics[width=0.25\textwidth]{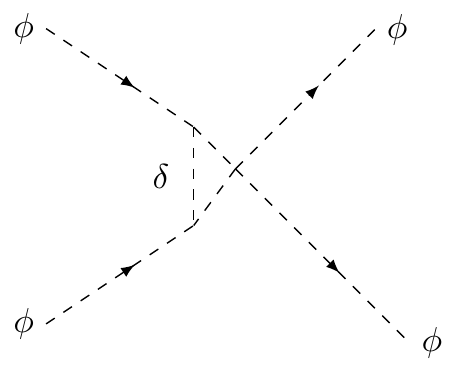}
\caption{\label{sigmam1_fig} Feynman diagrams for the processes contributing to the self-interaction of dark matter.}
 \end{center}
\end{figure}
In our model, the diagrams contributing to self-scattering are presented in the Fig.~\ref{sigmam1_fig}. The amplitude can be written as
\begin{eqnarray}
i\mathcal{M}&=& i\Big(\lambda_{11}+\mu_{\rm eff}^2\frac{1}{(s-m_\delta^2)}+\mu_{\rm eff}^2\frac{1}{(t-m_\delta^2)}+\mu_{\rm eff}^2\frac{1}{(u-m_\delta^2)}\Big),
\label{selfeq}
\end{eqnarray}
where $s$, $t$, $u$ are the Mandelstam variables and $\mu_{\rm eff} = \mu_{12} + \frac{\lambda_{12}}2 v_\delta$. The self-scattering cross-section thus becomes
\begin{eqnarray}
\sigma_{\rm{self}}&=&\frac{1}{64 \pi m_\phi^2}|\mathcal{M}|^2.
\end{eqnarray}

\begin{figure}[h!]
\begin{center}
\includegraphics[width=0.48\textwidth]{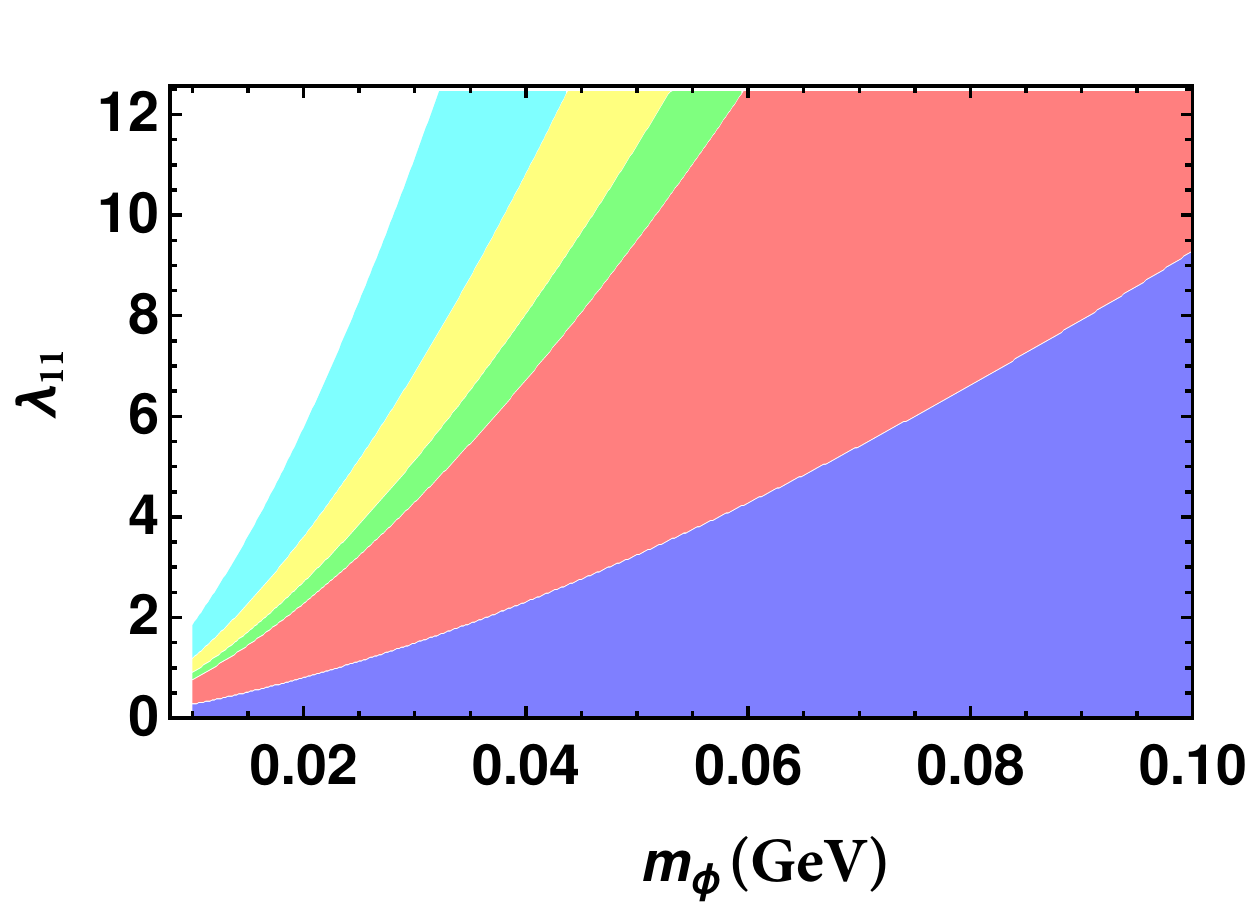} 
\includegraphics[width=0.48\textwidth]{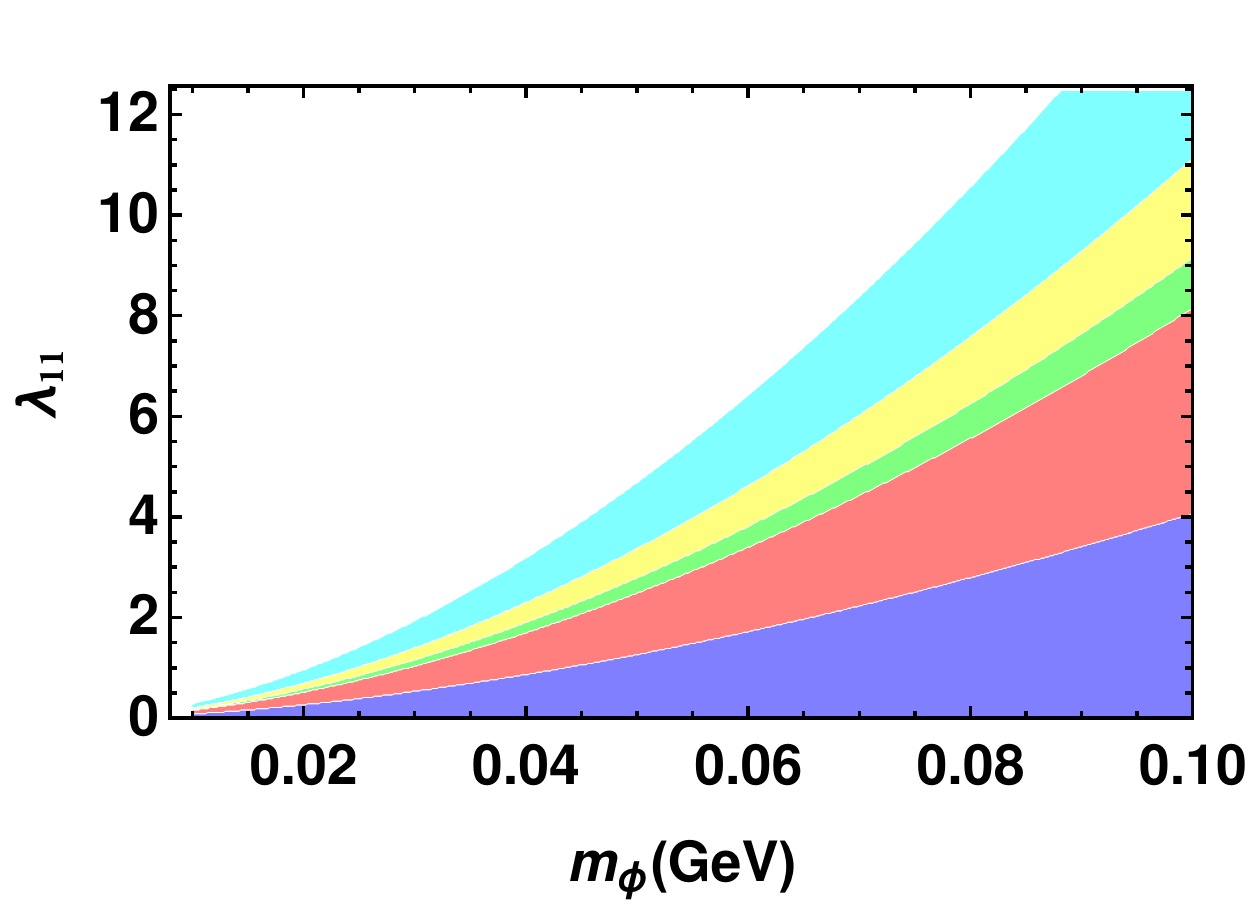} 
\caption{\label{sivev_fig} Regions showing the parameter space that satisfy $\sigma_{\rm self}/m_\phi\lesssim10~\rm{ cm^2/g}$. Color coding for various divisions in $\sigma_{\rm self}/m_\phi$ limit: $\sigma_{\rm self}/m_\phi\lesssim 0.1~\rm{ cm^2/g}$ (blue), $0.1~\rm{ cm^2/g}\lesssim\sigma_{\rm self}/m_\phi\lesssim 1~\rm{ cm^2/g}$ (red), $1~\rm{ cm^2/g}\lesssim\sigma_{\rm self}/m_\phi\lesssim 1.5~\rm{ cm^2/g}$ (green), $1.5~\rm{ cm^2/g}\lesssim\sigma_{\rm self}/m_\phi\lesssim 3~\rm{ cm^2/g}$ (yellow) and $3~\rm{ cm^2/g}\lesssim\sigma_{\rm self}/m_\phi\lesssim 10~\rm{ cm^2/g}$ (cyan).}
 \end{center}
\end{figure}

We compute the self scattering cross-section and analyze the bounds mentioned above. The first term in Eq.~\ref{selfeq} usually gives the dominant contribution to the self-scattering process and hence we study the allowed range of $\lambda_{11}$ as a function of the DM mass. Table~\ref{tab:sigmam} clearly shows that different observations suggest different limits on $\sigma_{\rm self}/m_\phi$ so it is not possible to choose a common range satisfying all of them. We have thus chosen an upper bound of $\sigma_{\rm self}/m_\phi\lesssim~ 10 ~\rm{ cm^2/g}$ in order to incorporate all the constraints in a single plot. In the Figure~\ref{sivev_fig} we have shown the allowed parameter space that satisfy the self-scattering bounds. We have divided the plots into various regions, i.e., (i) $\sigma_{\rm self}/m_\phi\lesssim 0.1~\rm{ cm^2/g}$ (blue), (ii) $0.1~\rm{ cm^2/g}\lesssim \sigma_{\rm self}/m_\phi\lesssim  1~\rm{ cm^2/g}$ (red), (iii) $1~\rm{ cm^2/g}\lesssim \sigma_{\rm self}/m_\phi\lesssim 1.5~\rm{ cm^2/g}$ (green), (iv) $1.5~\rm{ cm^2/g}\lesssim \sigma_{\rm self}/m_\phi\lesssim  3~\rm{ cm^2/g}$ (yellow) and (v) $3~\rm{ cm^2/g}\lesssim \sigma_{\rm self}/m_{\phi}\lesssim 10~\rm{ cm^2/g}$ (cyan). The left panel of Fig.~\ref{sivev_fig} corresponds to $v_\delta=10$ MeV while the right panel is for $v_\delta= 100$ MeV. As can be seen from the plots, an increase in the DM mass requires a larger value of $\lambda_{11}$ to satisfy the $\sigma_{\rm self}/m_{\phi}$ constraints (except for the blue region with $\sigma_{\rm self}/m_\phi\lesssim 0.1~\rm{ cm^2/g}$). This is quite expected as the DM self-scattering cross-section is proportional to its coupling strength. A larger value of $m_{\phi}$ requires a larger value of $\sigma_{\rm self}$ and hence a larger value of $\lambda_{11}$ in order to satisfy the lower bound on $\sigma_{\rm self}/m_\phi$. The absence of this feature in the blue region is because there is no lower limit on $\sigma_{\rm self}/m_\phi$ here. This dependence on $\lambda_{11}$ is more prominent for $v_\delta =10$ MeV (left panel) compared to $v_\delta=100$ MeV (right panel). For a low value of $v_\delta$, the dominant contribution to the self-scattering cross-section is from the contact interaction (four scalar interaction) diagram as the other process is suppressed by the propagator of $\delta$. As we increase $v_\delta$, the value of $\mu_{\rm eff}$ increases and the second term starts contributing and hence the dependence on $\lambda_{11}$ is flattened out a bit. This is clearly visible on the right panel of Fig.~\ref{sivev_fig} where $v_\delta = 100$ MeV. Another distinction between the two plots is that the allowed range of $\lambda_{11}$ for each region is much larger for the left panel plot compared to the right panel. As the second term in Eq.~\ref{selfeq} contributes significantly for the $v_\delta =100$ MeV case, a smaller value of $\lambda_{11}$ is required here. On the other hand, for $v_\delta=10$ MeV case, the entire self-interaction cross-section is due to the first term which is directly proportional to $\lambda_{11}^2$.  
\vspace*{-0.7cm}

\begin{figure}[h!]
\begin{center}
\includegraphics[width=0.48\textwidth]{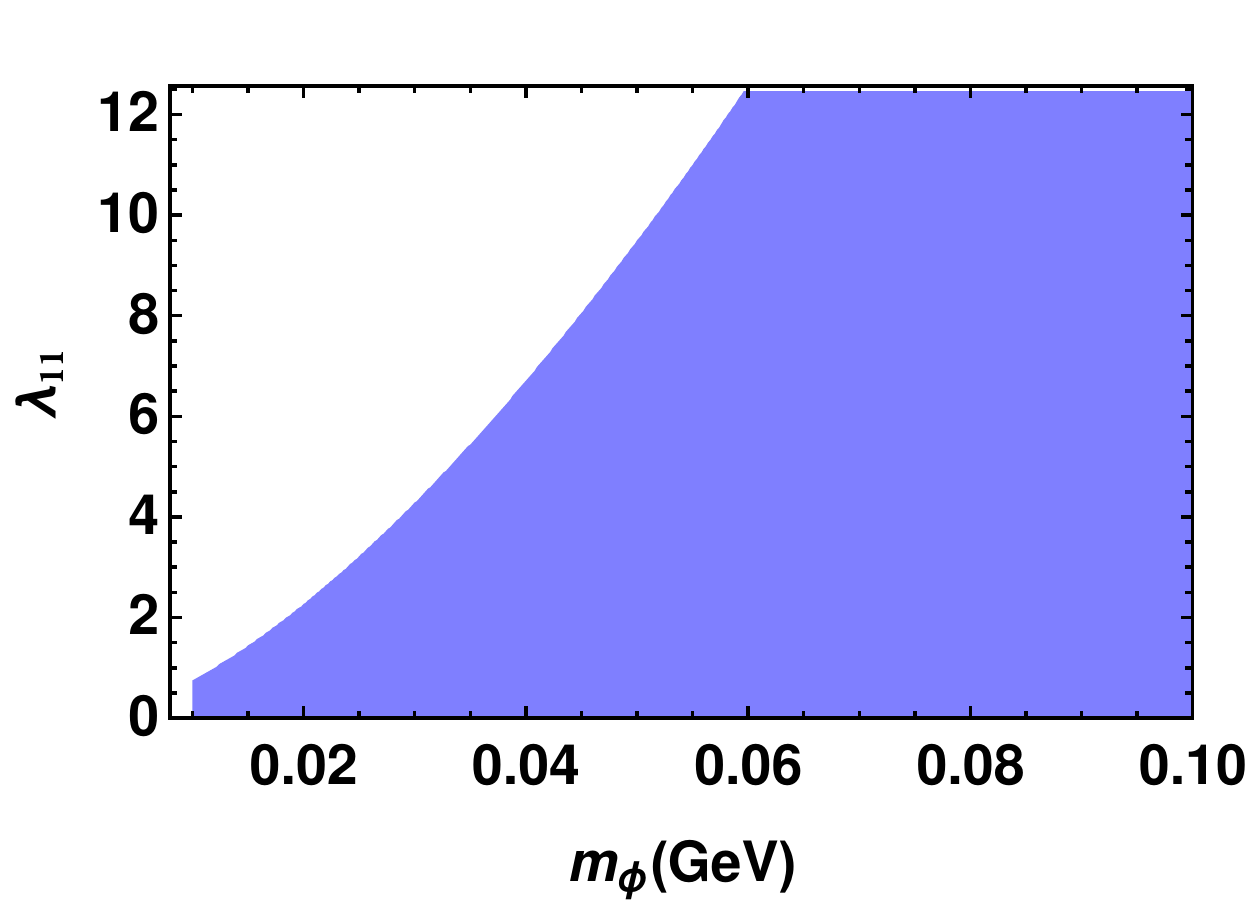} 
\includegraphics[width=0.48\textwidth]{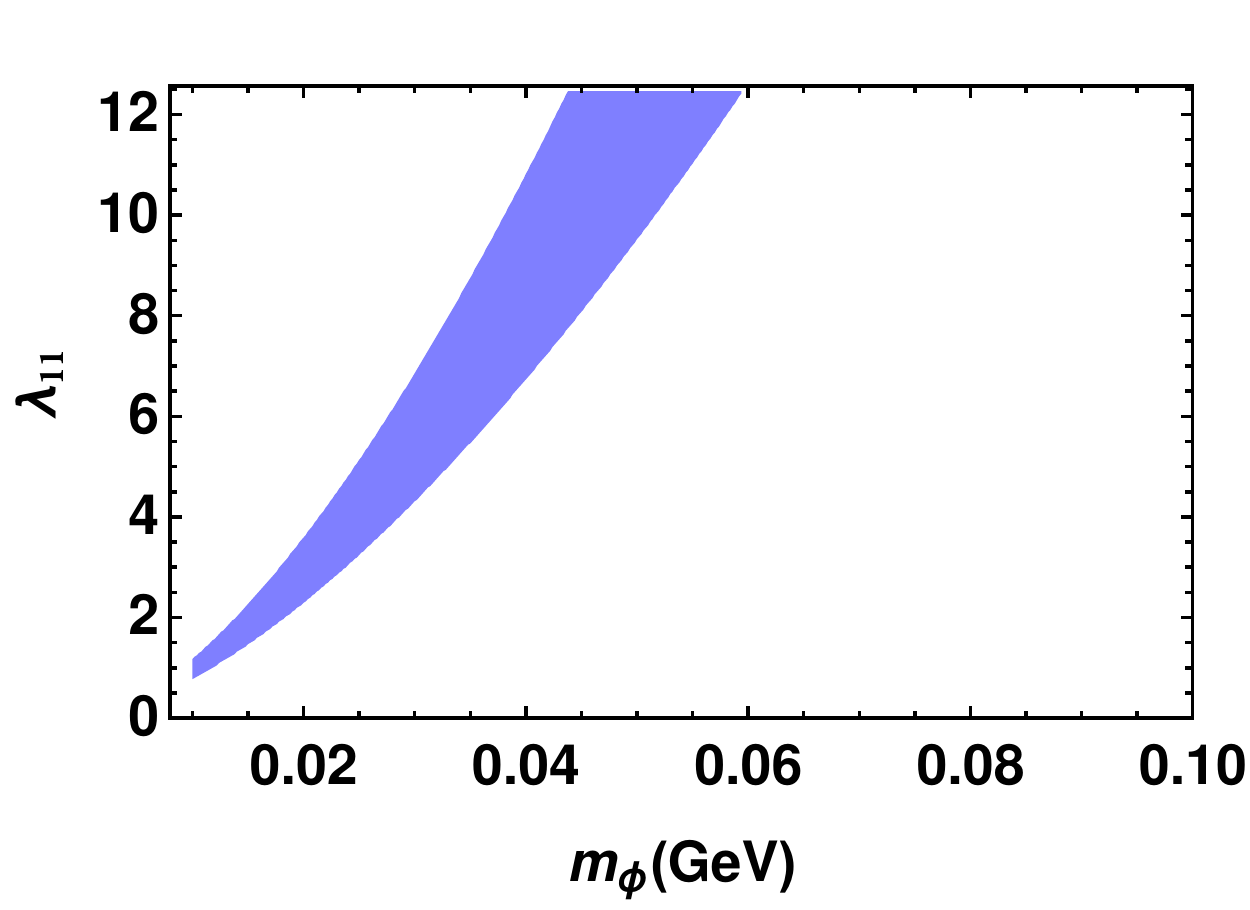} 
\caption{\label{sicluster_fig} Shaded regions in $\lambda_{11}$ and $m_\phi$ plane that satisfy the Bullet cluster (left) and Abell cluster data (right). VEV of $\delta$ is assumed to be 10 MeV. }
 \end{center}
\end{figure}

In Fig.~\ref{sicluster_fig} we consider the $\sigma_{\rm self}/{m_\phi}$ constraints arising from the recently observed Bullet and Abell galaxy clusters. The Bullet cluster bound suggests $\sigma_{\rm self}/m_\phi\lesssim~ 1~ \rm{ cm^2/g}$~\cite{Randall:2007ph} while Abell cluster suggest $1~\rm{ cm^2/g}\lesssim~\sigma_{\rm self}/m_\phi\lesssim~ 3~ \rm{ cm^2/g}$~\cite{10.1093/mnrasl/slv088}. The left panel in Fig.~\ref{sicluster_fig} represents the Bullet cluster bound and the right panel is for the Abell cluster bound for $v_\delta= 10$ MeV. As the two observations do not agree with each other, their preferred regions of allowed parameters are also completely disjointed in our model. It is thus not possible for us to choose an allowed range of parameters based on these two observations.


\section{Neutrino mass generation} \label{NM}

The inclusion of three right-handed neutrinos ($N_i$) in the model helps generate the light neutrino masses through type-I seesaw mechanism. The neutrino specific terms in the Lagrangian are given as
\begin{equation}
\mathcal{L}_Y \supset \left[ Y_{D_{ij}} L^T_i i \sigma_2 H N_j + H.C. \right] + M_{N_{ij}} N_i^c N_j + f_{ij} N_i^c N_j \delta.
\end{equation} 
The neutrino mass matrix is thus a $6\times6$ matrix which can be written as 
\begin{equation}
\begin{pmatrix}
0&M_D \\ M_D^T & M_R
\end{pmatrix},
\end{equation}
where $M_{D_{ij}}=Y_{D_{ij}} v_H$ is the Dirac mass term while $M_{R_{ij}} = M_{N_{ij}}+f_{ij} v_\delta$ is the Majorana mass term for the right-handed neutrino. Without loss of generality we can choose one of the right-handed neutrinos to be much lighter with mass around a few MeV. This will help in generating the required baryon asymmetry through leptogenesis as will be discussed in the next section. 

\begin{table}[t]

\begin{center}

{\renewcommand{\arraystretch}{1.5}

\begin{tabular}{||c|} \hline

 6.79$\times 10^{-5}~\text{eV}^2$ $<\Delta m_{21}^2<$ 8.01$\times 10^{-5}~\text{eV}^2$\\ \hline

 2.432$\times 10^{-3}~\text{eV}^2$ $<\Delta m_{31}^2<$ 2.618$\times 10^{-3}~\text{eV}^2$\\ \hline

 $0.275<\sin^2{\theta_{12}}<0.350$ \\ \hline

 $0.427<\sin^2{\theta_{23}}<0.609$ \\ \hline

 $0.02046<\sin^2{\theta_{13}}<0.02440$ \\ \hline

 $U_{PMNS}$ $ \begin{pmatrix}

0.797 \rightarrow 0.842 & 0.518 \rightarrow 0.585 & 0.143 \rightarrow 0.156 \\

0.244 \rightarrow 0.496 & 0.467 \rightarrow 0.678 & 0.646 \rightarrow 0.772 \\

0.287 \rightarrow 0.525 & 0.488 \rightarrow 0.693 & 0.618 \rightarrow 0.749\end{pmatrix} $ \\ \hline

\end{tabular}}

\caption{{Experimental $3\sigma$ ranges for light neutrino parameters from NuFIT~\cite{Esteban:2018azc, nufit41}.}}

\label{tab:numix}

\end{center}
\end{table}

\begin{figure}[h!]
\begin{center}
\includegraphics[width=0.49\textwidth]{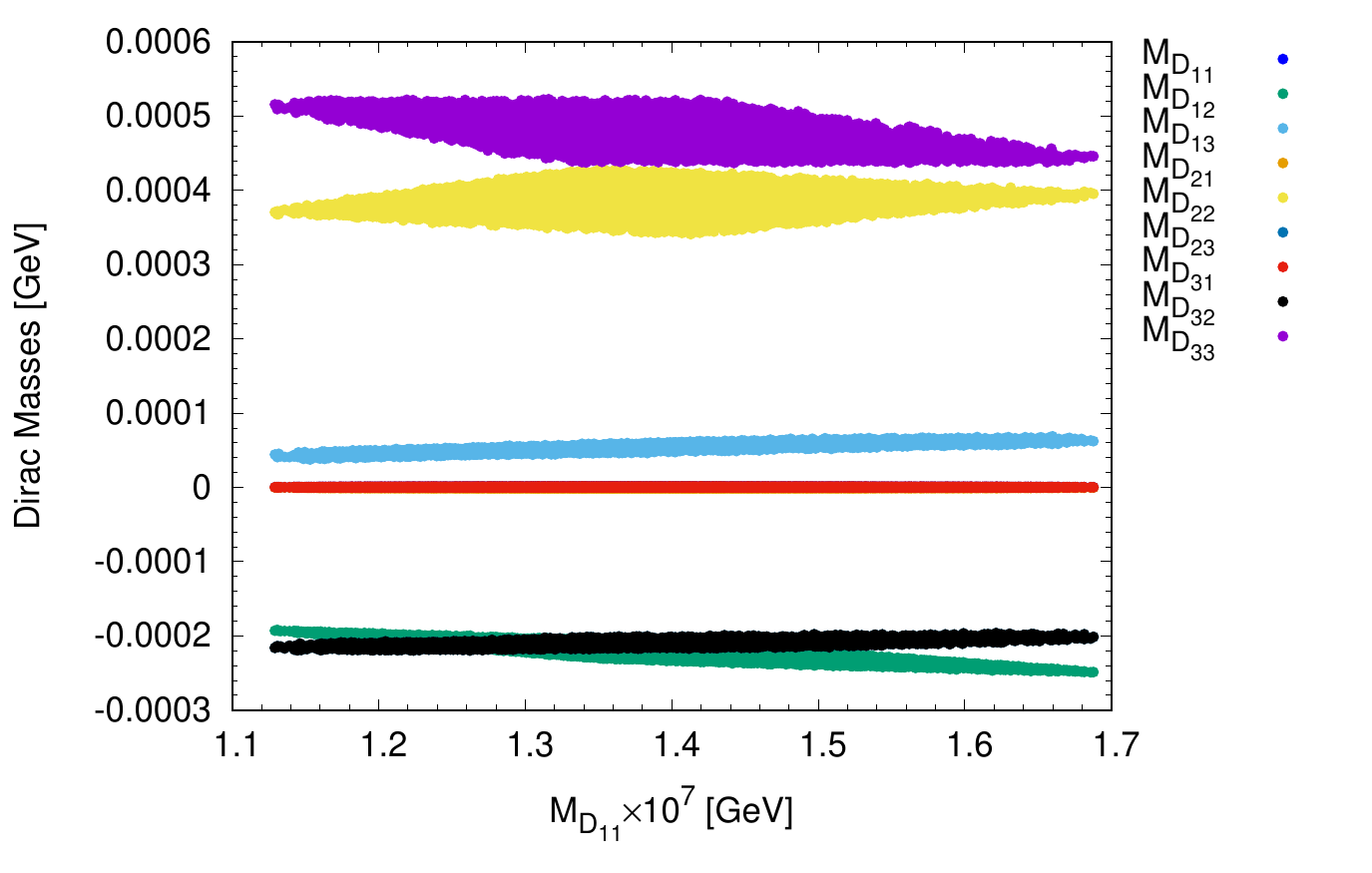} 
\includegraphics[width=0.49\textwidth]{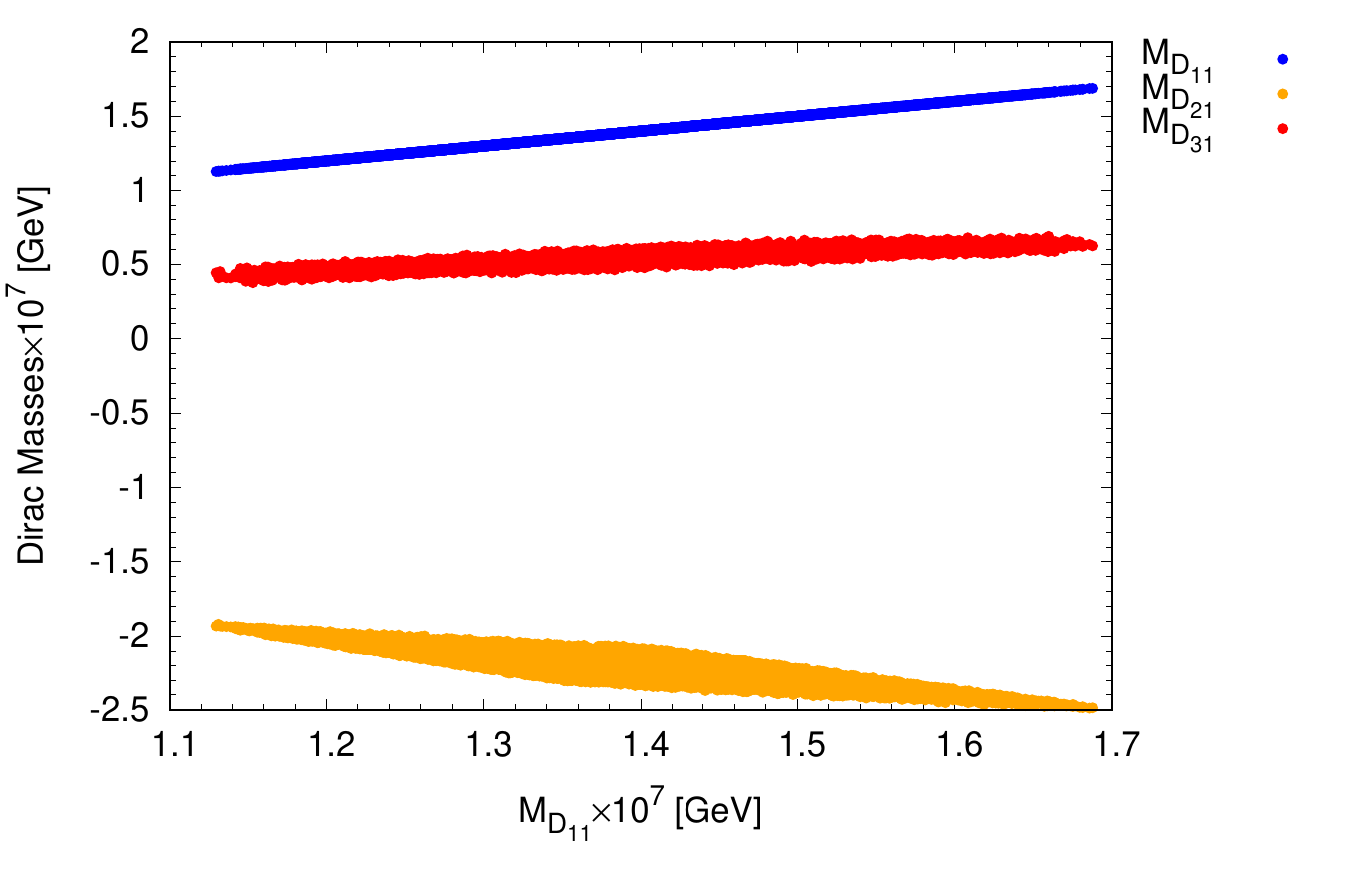} 
\caption{\label{fig:neutr} Neutrino Dirac masses satisfying the experimental mass-squared differences and mixings with heavy neutrino masses $M_{N_1}=10$~MeV, $M_{N_2} \approx 10 $~TeV, $M_{N_3} \approx 10 $~TeV.}
\end{center}
\end{figure}

The experimentally observed $3\sigma$ ranges \cite{Esteban:2018azc, nufit41} for the neutrino mass-squared differences and the corresponding Pontecorvo-Maki-Nakagawa-Sakata (PMNS) matrix elements for normal hierarchy case are given in Table~\ref{tab:numix}. In our scenario, we choose the two heavy right-handed neutrinos to be almost degenerate with a mass of around 10 TeV while the lighter right-handed neutrino mass is chosen to be around 10 MeV. For this mass spectrum of the singlet neutrinos, we perform a scan over the neutrino Dirac masses ($M_{D_{ij}}$) which can satisfy the experimental constraints. Fig.~\ref{fig:neutr} shows a scatter plot of the Dirac masses which satisfy the neutrino mass-squared differences and mixings as given in Table~\ref{tab:numix}.

It can be seen from Fig.~\ref{fig:neutr} that there is a hierarchy in the neutrino Dirac masses. As we have chosen the lightest singlet neutrino ($N_1$) to be only around 10 MeV, the mass terms $M_{D_{11}}$, $M_{D_{21}}$ and $M_{D_{31}}$ are much smaller than the other terms in order to satisfy the seesaw structure of the mass generation mechanism. As is clearly visible in the left panel of the figure, these three small masses appear almost degenerate compared to the other terms. To show the hierarchy between them, we have included the right panel which only shows these three smaller masses. For both the figures, we have plotted the Dirac masses with respect to the $M_{D_{11}}$ as it shows how the other mass terms are varying in magnitude as $M_{D_{11}}$ increases. 

The left panel of Fig.~\ref{fig:neutr} gives a clear indication that $M_{D_{33}}$ is the largest in each case. This can be understood as an effect of our choice of normal hierarchy for the neutrino mass spectra. Since $\nu_3$ is the heaviest, it implicitly requires the Dirac mass term $M_{D_{33}}$ to be the largest in each case. The plot also shows that if we neglect the much smaller $M_{D_{11}}$, $M_{D_{21}}$ and $M_{D_{31}}$ mass terms, the value of $M_{D_{13}}$ is the smallest among the rest. A similar feature can be observed in the right panel of Fig.~\ref{fig:neutr} where $M_{D_{31}}$ is the smallest. This is quite expected as the two terms $M_{D_{13}}$ and $M_{D_{31}}$ are responsible for generation of the $\theta_{13}$ mixing between $\nu_1$ and $\nu_3$. The smallness of the experimentally measured value of $\theta_{13}$ thus requires these two Dirac mass terms to be quite small as well. 

Another interesting feature in this figure is that the magnitude of $M_{D_{12}}$ and $M_{D_{21}}$ increases as $M_{D_{11}}$ increases. Since the $M_{D_{12}}$ and $M_{D_{21}}$ are responsible for generation of the mixing between $\nu_1$ and $\nu_2$, it is quite obvious that these term have to increase as the diagonal terms increase. This feature can also be seen in $M_{D_{13}}$ and $M_{D_{31}}$ mass terms though it is not as pronounced as the other case, maybe due to the smallness of $\theta_{13}$.

\section{Leptogenesis} \label{LEP}

The observed baryon asymmetry of the universe can be explained in this model through leptogenesis. The CP asymmetry is generated through the interference of the tree-level and one-loop diagrams for the heavy neutrino decay. Let us consider the decay of one of the heavy right-handed neutrinos into a final state lepton and a scalar boson. There are two possible decays here. The heavy neutrino ($N_2$ or $N_3$) can decay into an active neutrino and Higgs boson or it can decay into the lightest right-handed neutrino ($N_1$) and singlet scalar boson ($\delta$).
\begin{figure}[h!]
\begin{center}
\includegraphics[width=0.8\textwidth]{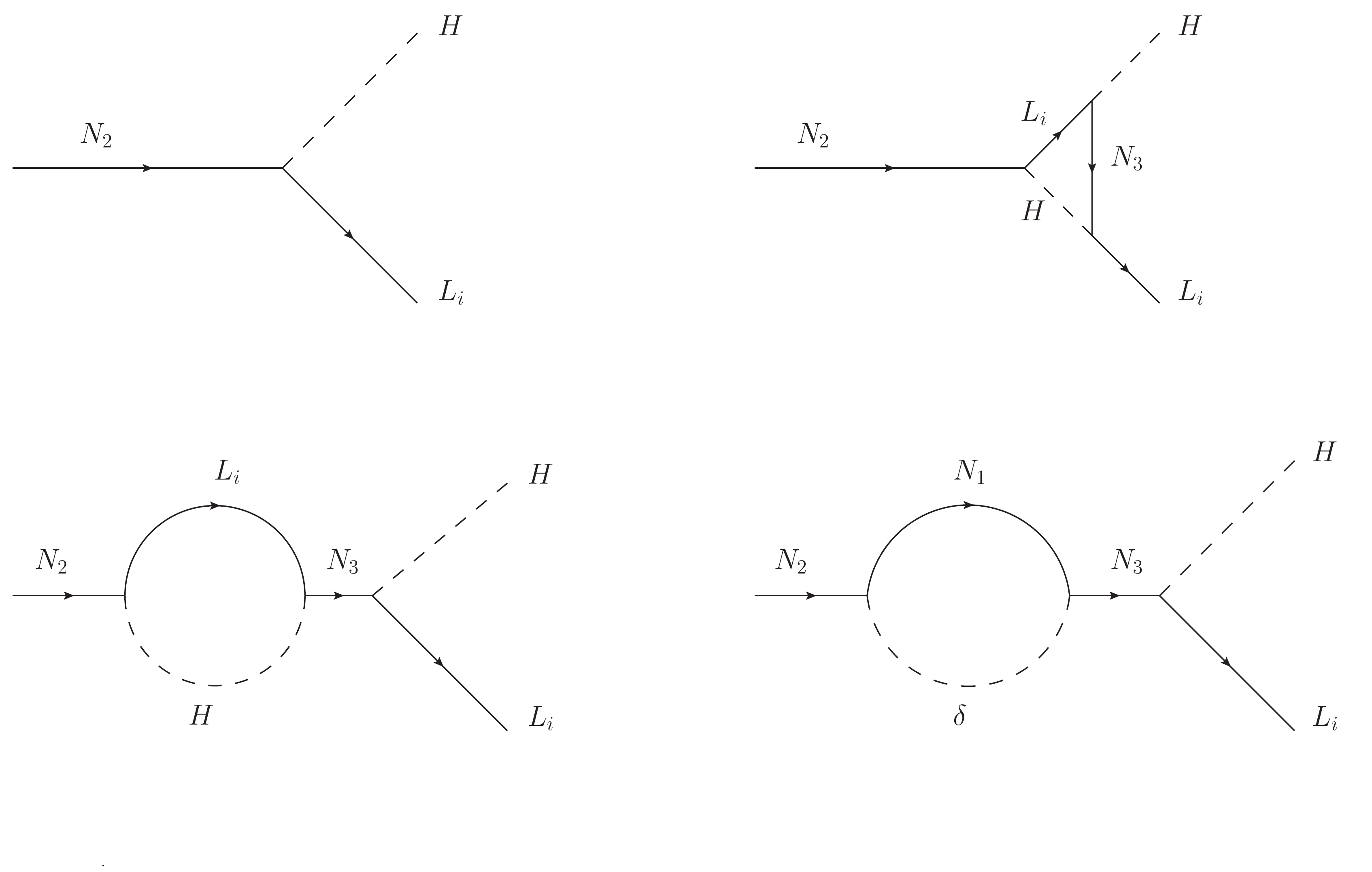} 
\caption{\label{fig:lepto} Diagrams contributing to leptogenesis. The required CP asymmetry is generated through the interference of the tree-level process with the one-loop diagrams.}
\end{center}
\end{figure}
Assuming the final state leptons and scalar boson masses are much smaller compared to the right-handed neutrino, the required lepton number asymmetry will be given as
\begin{eqnarray}
\epsilon_{1} &=& -\sum_i\left[\frac{\Gamma(N_2
\to \bar{l_i}H^{\ast}) - \Gamma(N_2 \to l_i H) }{\Gamma_{\rm
tot}(N_2)}\right].
\end{eqnarray}
Here $N_2$ represents the heavy right-handed neutrino and $\Gamma_{\rm tot}(N_2)$
is its total decay width given as  
\begin{equation}
\label{eq:vv}
\Gamma_{\rm tot}(N_2)={(Y_{D_{2i}}^\dagger Y_{D_{2i}})+|f_{12}|^2
\over 4\pi}M_{N_2}.
\end{equation}
The $Y_D$ term corresponds to the decay into a SM Higgs boson and a light neutrino while the $f$ coupling is due to the decay into a light singlet neutrino and singlet scalar boson. The one-loop diagrams for these decay processes will involve the generic self-energy and vertex corrections due to the $Y_D$ term in the Yukawa Lagrangian. In addition, there will be an extra self-energy correction due to the $fN_1^c N_j \delta$ term as well. Fig.~\ref{fig:lepto} shows the tree level and one-loop Feynman diagrams contributing to the process. As a result, the CP asymmetry in this case will be given as 
\begin{equation}
\epsilon_1 = \frac{1}{8\pi} \left( [ g_V(x)+
g_S(x)]{\cal T}_{23} + g_S(x){\cal S}_{23}\right),
\label{eq:asymm}
\end{equation}
where $g_V(x)=\sqrt{x}\{1-(1+x) {\rm ln}[(1+x)/x]\}$,
$g_S(x)=\sqrt{x}/(1-x) $ with $x=M_{N_3}^2/M_{N_2}^2$,
\begin{equation}
  \label{eq:epsilon}
{\cal T}_{23}={{\rm Im}[(Y_{D_{2i}} Y_{D_{3i}}^\dagger)^2]
 \over (Y_{D_{2i}}^\dagger Y_{D_{2i}}) +|f_{21}|^2}
\end{equation}
and
\begin{equation}
{\cal S}_{k1}={{\rm Im}[(Y_{D_{2i}} Y_{D_{3i}}^\dagger)(f_{21}
f_{31}^\dagger)]
 \over (Y_{D_{2i}}^\dagger Y_{D_{2i}}) +|f_{21}|^2}.
\end{equation}
Here the $g_V(x)$ term arrives from the vertex correction while the $g_S(x)$ terms are due to the self energy corrections. The ${\cal S}_{k1}$ term is purely due to the new contribution from the self energy diagram involving the singlet scalar field as discussed before.

It is quite interesting to see that when $M_{N_2}\simeq M_{N_3}$ (i.e. $x \simeq 1$), the correction from the self energy term is dominant and can significantly enhance the asymmetry factor given in Eq.~\ref{eq:asymm}. The lepton asymmetry in this case can be approximately given by
 \begin{eqnarray}
 \epsilon_1 & \simeq & -\frac{1}{16\pi}\left[
           \frac{M_{N_3}}{v^2}\frac{Im[(Y^{\ast}_D m_{\nu}Y^{\dagger}_D)_{22}]}
           {(Y_D^\dagger Y_D)_{22}+|f_{21}|^2}
        +\frac{ Im[(Y_DY^{\dagger}_D)_{23}(f_{21}f_{31}^{\dagger})]}
                   {(Y_D^\dagger Y_D)_{22}+|f_{21}|^2}\right]R~,
\label{eq:epsilon2}
\end{eqnarray}
where $R$ may be recognized as a resonance factor defined as $R \equiv |M_{N_2}|/(|M_{N_3}|-|M_{N_2}|)$. In the absence of the $f_{ij}$ couplings, only the first term of Eq.~\ref{eq:epsilon2} will contribute and a value of $R\sim 10^{6-7}$ is required to get the correct amount of lepton number asymmetry. This leads to almost degenerate masses for the heavy right-handed neutrinos with $M_{N_3} - M_{N_2} \approx 10^{-2}$ GeV resulting in an unnaturally large fine-tuned scenario. The presence of the extra couplings between the singlet scalar and singlet neutrinos can alleviate this problem by enhancing the second term in Eq.~\ref{eq:epsilon2}. The $f_{21}$ coupling though is constrained from the the out-of-equilibrium condition which requires the total decay width of $N_2$ to be less than the Hubble constant. Thus we get
\begin{equation}
\Gamma_{N_2} < H|_{T=M_{N_2}}
\end{equation}
where $H$ is the Hubble expansion rate. This produces an upper bound on $f_{21}$ with
\begin{eqnarray}
\sqrt{|f_{21}|^2}<3\times 10^{-4}\sqrt{M_{N_2}/10^9(\mbox{GeV})}.
\end{eqnarray}
This constraint however is not applicable for the $f_{31}$ coupling and it can be taken to be much larger. The $B-L$ asymmetry generated in this scenario is given as 
\begin{equation}
Y_{B-L} = - \eta \epsilon_1 Y^{eq}_{N_2},
\end{equation}
where $\eta$ is the efficiency factor and $Y^{eq}_{N_2} \simeq \frac{45}{\pi^4}\frac{\zeta(3)}{g_{\ast}\,k_B} \frac{3}{4}$ is the number density of $N_2$ in thermal equilibrium at $T\!>\!\!>\!M_{N_1}$. Here $k_B$ is the Boltzmann constant and $g_{\ast}$ is the effective number of degrees of freedom. This asymmetry can then be transferred into a baryon asymmetry through sphaleron induced processes at the electroweak symmetry breaking scale.

If one chooses $f_{21} \sim Y_{D_{21}}$ with $ f_{31} \sim \chi Y_{D_{31}}$, then the second term in Eq.~\ref{eq:epsilon2} increases by a factor of $\chi$ and the required asymmetry can be obtained even with a much smaller resonance factor $R$. If we choose $\chi \sim 10^3$, the value of $R$ required is only around $10^{3-4}$ and even a mass difference of a few GeV between the two heavy neutrinos could be allowed in order to generate the $B-L$ asymmetry. Thus we arrive at a scenario which is free from the extreme fine-tuning between the heavy neutrino masses and is instead realized through a simple hierarchy in the $f_{ij}$ couplings which is much more natural. 

\section{Neutrinoless Double Beta Decay} \label{N2BD}

\vspace*{-2pt}
\begin{figure}[h!]
\begin{center}
\includegraphics[width=0.3 \textwidth]{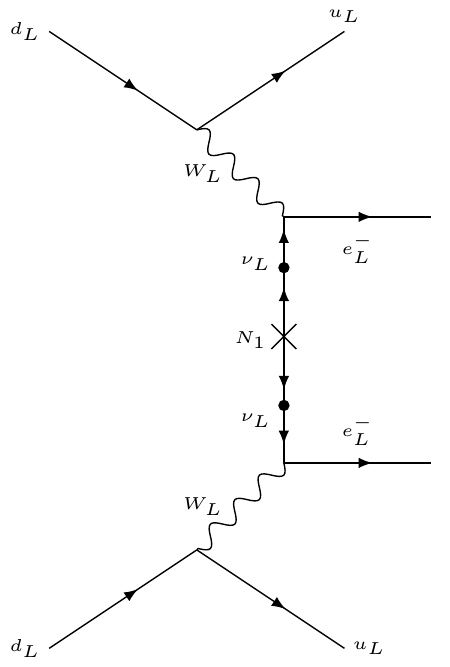}  
\caption{\label{fig:0vbb} Feynman diagram for dominant contribution to the neutrinoless double beta decay process mediated by the light right-handed neutrino in this model.}
\end{center}
\end{figure}
\vspace*{-2pt}

The low mass of $N_1$ can results in significant bound from neutrinoless double beta decay ($0\nu 2\beta$) processes \cite{Rodejohann:2011mu, Deppisch:2012nb, Graf:2018ozy}. The Feynman diagram in Fig.~\ref{fig:0vbb} shows the dominant contribution to $0\nu 2\beta$ signal in this model. As is quite evident from the diagram, the $0\nu 2\beta$ decay process will be directly proportional to the mass of the light right-handed neutrino and its mixing with the electron neutrino. This process can thus produce significant constraints on the allowed mixing between the active and sterile neutrinos. A detailed calculation of this diagram and its effect on the neutrino spectrum will be discussed in an upcoming paper.

\section{Conclusion} \label{CON}

In this paper we have constructed a unified model of a singlet scalar self-interacting DM with a mass of $\sim$ 10 MeV. We have given a $3 \to 2$ annihilation mechanism for generating the required relic density of the DM with a second singlet scalar in the final state. This second singlet acquires a non-zero VEV which helps in generating the DM mass. Solving the coupled Boltzmann equations, we identify the allowed region of parameter space that satisfy the experimentally observed bounds on DM relic abundance. Our results show that a larger DM mass requires a larger value of the self-interaction coupling $\lambda_{11}$ to satisfy the DM relic density. This is because the $3 \to 2$ annihilation process is mediated by the DM particle and hence is inversely proportional to its mass. One thus needs to increase the DM self-interaction coupling. A similar result can be obtained by simply increasing the scalar VEV while keeping $\lambda_{11}$ constant as this also helps in enhancing the DM annihilation cross-section. Self-interacting DM can have stringent bounds from several astrophysical observations. We analyze the model for bounds from many such observations focusing specifically on the recently observed Bullet and Abell galaxy clusters. Since these two bounds are mutually exclusive, we have presented our allowed parameter regions which can satisfy the constraints from either of them. 

We have also included three right-handed neutrinos which help generate the light neutrino masses via Type-I seesaw mechanism. We choose two of the right-handed neutrinos to be heavy around $10$ TeV while one of them remains light at around $10$ MeV. Detailed scans of the neutrino parameters were performed to identify a distinct pattern in their Dirac mass terms which is required to satisfy the experimental observations. This model can also explain the observed baryon asymmetry of the universe. The interaction term of the right-handed neutrinos with the scalar singlet can help generate a large lepton asymmetry which can then be converted into baryon asymmetry through sphaleron transitions. One of the possible way to test this model is through neutrino-less double beta decay as the light right-handed neutrino can have significant contribution to the process. The absence of any such signals will put stringent bounds on the model.

\bibliography{simpnu}
\bibliographystyle{JHEP}

%
%
\end{document}